\documentclass{ceurart}

\usepackage{amsmath,amssymb}
\usepackage{booktabs}
\usepackage{graphicx}
\usepackage{tikz}
\usepackage{pgfplots}
\pgfplotsset{compat=1.18}
\usetikzlibrary{shapes.geometric, arrows.meta, positioning, fit, calc, patterns}
\usepackage{multirow}
\usepackage[table]{xcolor}
\usepackage{colortbl}
\usepackage{enumitem}
\usepackage{subcaption}

\begin{document}

\copyrightyear{2026}
\copyrightclause{Copyright for this paper by its authors.
  Use permitted under Creative Commons License Attribution 4.0
  International (CC BY 4.0).}
\conference{ECOM'26: SIGIR Workshop on eCommerce, Jul 24, 2026, Melbourne, Australia}

\title{Scaling Dense Retrieval with LLM-Annotated Training Data: Structured Mining and Progressive Curriculum for E-Commerce Sponsored Search}

\author[1]{Md Omar Faruk Rokon}[email=mdomarfaruk.rokon@walmart.com]
\cormark[1]
\author[1]{Shasvat Desai}[email=shasvat.desai@walmart.com]
\cormark[1]
\author[1]{Jhalak Nilesh Acharya}[email=jhalak.acharya@walmart.com]
\cormark[1]
\author[1]{Isha Shah}[email=isha.shah@walmart.com]
\author[1]{Kumar Priyam}[email=kumar.priyam@walmart.com]
\author[1]{Brahanyaa Somasundaram}[email=brahanyaa.somasundaram@walmart.com]
\author[1]{Vamsee Tangirala}[email=vamsee.tangirala@walmart.com]
\author[1]{Minuteresa Thomas}[email=minuteresa.thomas@walmart.com]
\author[1]{Vivek Arora}[email=vivek.arora@walmart.com]
\author[1]{Vijay Manchi}[email=vijay.manchi@walmart.com]
\author[1]{Hong Yao}[email=hong.yao0@walmart.com]
\author[1]{Kuang-chih Lee}[email=kuang-chih.lee@walmart.com]

\address[1]{Walmart Global Tech, Sunnyvale, CA, USA}

\cortext[1]{Corresponding author.}

\begin{abstract}
How can we generate high-quality training data for dense retrieval models at production scale, without relying on click signals or manual annotation? This question is critical for e-commerce sponsored search, where click-based training suffers from position bias and tail-query sparsity, and manual labeling at the scale of hundreds of millions of query-item pairs is economically infeasible. Our work is driven by the following insight: heterogeneous retrieval systems disagree on most items they retrieve, and this disagreement creates a natural source of structured training signal---easy positives where all systems agree, hard positives that only lexical systems find, and hard negatives that fool exactly one system. As our key novelty, we combine three ideas into an end-to-end pipeline: (a) multi-channel retrieval mining with rank metadata from three production systems, (b) graded-relevance annotation by a calibrated three-model cascade (184M cross-encoder $\to$ LoRA-adapted 2B LLM $\to$ LoRA-adapted 8B LLM) that reaches 89.1\% agreement with trained human annotators, and (c) three-stage progressive curriculum training (BCE $\to$ MNR $\to$ Triplet) that organizes 240M+ training examples across five difficulty levels. We deploy the trained two-tower BERT model on Walmart's sponsored search and evaluate it against 30K queries labeled by trained third-party human annotators. First, we show that the system achieves +5.1\% NDCG@10 (0.878 $\to$ 0.923) over the click-trained production baseline, with the largest gain on tail queries (+6.8\%). Second, we show that embarrassing retrievals (rating 0) drop from 8.7\% to 3.5\%. Third, a two-week online A/B test with tens of millions of ad requests per arm confirms +2.80\% ad spend, +1.4\% CTR, +2.8\% eCPM, and +2.9\% click conversion rate. Overall, our work provides a practical and scalable blueprint for replacing click-based training with structured LLM-annotated supervision in production retrieval systems.
\end{abstract}


\maketitle

\section{Introduction}
\label{sec:intro}

How can we train dense retrieval models for e-commerce sponsored search at production scale, when click signals are biased and manual annotation is too expensive? This question is important because retrieval is the first stage in a sponsored search pipeline: it determines which ads a user sees, and everything downstream---ranking, bidding, revenue---depends on its quality~\cite{aiello2016relevance, wang2023clickconv, wang2024semantic}. On platforms like Walmart.com, even small improvements in retrieval relevance translate to substantial business outcomes.

Consider a concrete example. When a user searches for ``iPhone charger,'' the retrieval model must surface relevant sponsored items such as Lightning cables and USB-C adapters, while avoiding embarrassing results like phone cases or screen protectors. A click-trained model learns which items users clicked on for this query in the past---but clicks are noisy. Users click top-ranked results regardless of relevance (position bias~\cite{wang2023clickconv, su2018user}), and for tail queries like ``MagSafe charger for iPhone 15 Pro Max,'' there may be no click history at all (the cold-start problem). Manual annotation can produce clean labels, but at the scale we need---240M+ query-item pairs across 4M queries---it would cost tens of millions of dollars per refresh cycle, which is infeasible for weekly model retraining.

Our work is driven by the following insight: the three retrieval systems already running in production---a dictionary-based lexical system, a BM25 system, and the current ANN embedding model---disagree on most of the items they retrieve. At the top 500, their pairwise overlap is only 13--15\%. This disagreement is not noise; it is structured signal. Items that all three systems retrieve and that an LLM judges relevant are high-confidence easy positives. Items that lexical systems find but the ANN misses are hard positives---exactly the failures we want the next model to fix. Items retrieved by only one system and judged irrelevant are hard negatives that confuse real production systems. By combining multi-channel disagreement with a calibrated LLM annotation cascade, we can mine millions of structured training examples at five distinct difficulty levels, without any click signals or manual labels.

As our key novelty, we combine three ideas into a single end-to-end pipeline: (a) structured data mining from multi-channel retrieval disagreement, using rank position, channel agreement, and catalog-level token similarity to create five difficulty levels of training examples; (b) a calibrated annotation cascade of three fine-tuned classifiers (184M cross-encoder $\to$ LoRA-adapted 2B LLM $\to$ LoRA-adapted 8B LLM) with per-class isotonic calibration, reaching 89.1\% agreement with trained human annotators---extending our prior work on single-model binary annotation~\cite{rokon2024enhancement} to a multi-model cascade on 5-class graded relevance; and (c) progressive curriculum training across three stages of increasing difficulty (BCE $\to$ MNR $\to$ Triplet), where each stage uses a loss function matched to the discrimination granularity required by its sample types. The pipeline generates 240M+ training examples across 4M queries and enables fully automated weekly model retraining. The approach is a form of distant supervision~\cite{mintz2009distant}: the cascade is used once, offline, to produce labels; the student is then trained with standard supervised objectives. We intentionally do not perform classical knowledge distillation~\cite{hinton2015distilling}---no soft-label transfer, no teacher gradient flow, no co-training---because decoupling the annotator from the student means annotations can be cached, reused, and the two components upgraded independently.

We deploy the resulting model (Embedding Model V3) on Walmart's sponsored search and evaluate it against 30K queries labeled by trained third-party human annotators---independent of the cascade used for training. First, we show that the system achieves +5.1\% NDCG@10 (0.878 $\to$ 0.923) over the click-trained production baseline, with embarrassing results reduced from 8.7\% to 3.5\%. Second, we show that improvements are consistent across all query segments, with the largest gain on tail queries (+6.8\% NDCG@10), consistent with addressing the cold-start weakness of click-trained models. Third, a two-week online A/B test with tens of millions of ad requests per arm suggests that these offline gains transfer to online business metrics: +2.80\% ad spend, +1.4\% CTR, +2.8\% eCPM, and +2.9\% click conversion rate.

Our contributions are: (a) a structured data mining pipeline that converts multi-channel retrieval disagreement into five difficulty levels of training examples---the key upstream innovation that lets a standard two-tower BERT achieve production-quality retrieval without architectural changes; (b) a calibrated three-model annotation cascade with per-class isotonic calibration that halves compute relative to running all models on every pair, while maintaining 89.1\% agreement with human annotators on a 5-class graded-relevance task; (c) a three-stage progressive curriculum (BCE $\to$ MNR $\to$ Triplet) that yields +9.5\% NDCG@10 over single-stage training on the same data; and (d) production deployment and online A/B validation at Walmart scale, providing evidence that offline relevance gains transfer to measurable business outcomes.

The remainder of this paper is organized as follows. Section~\ref{sec:related} reviews related work on neural retrieval, training data generation, and LLM-based annotation. Section~\ref{sec:problem} defines the problem and the training data challenge. Section~\ref{sec:method} describes our four-stage pipeline in detail. Section~\ref{sec:experiments} presents the experimental setup, and Section~\ref{sec:results} reports results. Section~\ref{sec:discussion} discusses findings and limitations, and Section~\ref{sec:conclusion} concludes.

\section{Related Work}
\label{sec:related}

Our work sits at the intersection of four research areas: neural retrieval architectures, training data generation, LLM-based annotation, and curriculum learning.

\textbf{Neural retrieval for e-commerce.} The two-tower architecture was introduced by Huang et al.~\cite{huang2013learning} as DSSM and later adopted for YouTube recommendations~\cite{covington2016deep}. It has since become the dominant paradigm for large-scale retrieval~\cite{karpukhin2020dense}. In e-commerce, Nigam et al.~\cite{nigam2019semantic} built semantic product search at Amazon, Huang et al.~\cite{huang2020embedding} deployed embedding-based retrieval at Facebook, and Fan et al.~\cite{fan2019mobius} proposed MOBIUS for Baidu's sponsored search. At Walmart, Magnani et al.~\cite{magnani2022semantic} introduced semantic retrieval for product search, Wang et al.~\cite{wang2024semantic} proposed progressive fusion training for ads retrieval, and Desai et al.~\cite{desai2026unified} created a unified supervision framework using relevance and engagement signals. Our work does not propose a new model architecture. Instead, we address the upstream problem of training data quality, which is orthogonal to these architectural advances.

\textbf{General-purpose dense retrievers.} Recent work has produced strong general-purpose embedding models such as E5~\cite{wang2022e5}, BGE~\cite{xiao2023cpack}, and GTE~\cite{li2023gte}, which achieve impressive zero-shot performance on the BEIR benchmark~\cite{thakur2021beir}. Learned sparse approaches like SPLADE~\cite{formal2021splade} offer a complementary efficiency profile. These models are strong off-the-shelf baselines, but our production setting imposes constraints they do not address out of the box: sponsored-ad catalog matching, strict serving latency requirements with ANN indexing, and weekly refresh on a shifting query distribution. This is why we train a domain-adapted student rather than serving a frozen general retriever.

\textbf{Distillation and distant supervision for retrieval.} Knowledge distillation~\cite{hinton2015distilling} transfers a teacher's soft label distribution to a smaller student. In retrieval, Hofst{\"a}tter et al.~\cite{hofstatter2020improving} proposed cross-architecture distillation for efficient ranking, Qu et al.~\cite{qu2021rocketqa} introduced RocketQA with cross-encoder supervision, and Reimers and Gurevych~\cite{reimers2020making} distilled monolingual embeddings into multilingual models. All of these methods require the teacher during student training. Our approach is different: it is better described as distant supervision~\cite{mintz2009distant}. The teacher (LLM cascade) is used once, offline, to produce hard labels, and the student is trained with standard supervised objectives. This is conceptually similar to weakly-supervised pre-training~\cite{wang2022e5, xiao2023cpack} and LLM-generated training data~\cite{bonifacio2022inpars, dai2023promptagator}, but differs in its source---we mine real production query-item pairs rather than generating synthetic queries---and in its scale and production integration.

\textbf{Training data generation for retrieval.} Generating high-quality training data at scale is a long-standing challenge. Bonifacio et al.~\cite{bonifacio2022inpars} used LLMs for synthetic query generation, and Dai et al.~\cite{dai2023promptagator} proposed Promptagator for few-shot dense retrieval via LLM-generated queries. Reddy et al.~\cite{reddy2022shopping} released the Shopping Queries Dataset (ESCI) with multi-class relevance labels for e-commerce retrieval evaluation. Unlike these approaches, we do not generate synthetic data. Instead, we mine real production query-item pairs from multiple retrieval channels and annotate them with calibrated LLM labels, producing training data that directly reflects the production query distribution.

\textbf{LLMs as relevance judges.} Recent work shows that LLMs can serve as effective relevance judges. Thomas et al.~\cite{thomas2024large} showed that LLMs accurately predict searcher preferences, Faggioli et al.~\cite{faggioli2023perspectives} surveyed LLMs for IR relevance judgment, and Zheng et al.~\cite{zheng2023llmjudge} characterized the biases and strengths of LLM-as-judge protocols. In e-commerce, Rokon et al.~\cite{rokon2024enhancement} showed that a single LoRA-adapted LLM reaches 89.43\% accuracy on binary query-ad relevance classification. Our work extends that line in three ways: (1) we move from binary to 5-class graded relevance, which is harder and more useful for ranking; (2) we introduce a three-model cascade with per-class isotonic calibration~\cite{chen2023frugalgpt} that halves compute at equivalent accuracy; and (3) we integrate this cascade into a larger pipeline with multi-channel mining, curriculum training, and production deployment.

\textbf{Hard negative mining and curriculum learning.} The quality of negative examples is critical for contrastive training~\cite{xiong2021approximate, zhan2021optimizing, robinson2021contrastive}. ANCE~\cite{xiong2021approximate} proposed asynchronous index-based hard negative sampling, and Zhan et al.~\cite{zhan2021optimizing} studied optimal hard negative selection strategies. Curriculum learning~\cite{bengio2009curriculum} advocates organizing training data from easy to hard. Our approach combines both: we mine multiple difficulty levels of negatives from multi-channel disagreement patterns and train across three stages of increasing difficulty, each with a distinct loss function matched to the sample type.

\section{Problem Formulation}
\label{sec:problem}

\subsection{Task Definition}

We focus on the retrieval stage of sponsored search. Given a user search query $q \in \mathcal{Q}$, the goal is to identify a set of relevant sponsored items $\mathcal{I}_q \subset \mathcal{C}$ from the product catalog $\mathcal{C}$, such that items in $\mathcal{I}_q$ are relevant to $q$ and have active advertiser bids. We measure relevance on a five-point graded scale~\cite{jarvelin2002cumulated}:
\begin{equation}
    \text{rel}: \mathcal{Q} \times \mathcal{C} \rightarrow \{0, 1, 2, 3, 4\}
    \label{eq:relevance}
\end{equation}
where 0 = Embarrassing, 1 = Bad, 2 = Okay, 3 = Good, and 4 = Excellent. Items with $\text{rel}(q, i) \geq 3$ are considered relevant for retrieval purposes. The retrieval model must serve results within strict latency constraints suitable for real-time search.

\subsection{The Training Data Challenge}

The core challenge is generating training data at the scale and quality needed for weekly model retraining. There are three bottlenecks.

\textbf{Clicks are biased.} The probability of a click depends on display position, not just relevance: $P(\text{click} \mid q, i, \text{pos}) \neq P(\text{rel} \mid q, i)$. Users click top-ranked results regardless of quality~\cite{wang2023clickconv, su2018user}, so models trained on clicks learn to replicate the existing system's ranking rather than true relevance.

\textbf{Clicks are sparse.} For tail queries, clicks approach zero: $\mathbb{E}[\text{clicks}(q, i)] \to 0$ for $q \in \mathcal{Q}_{\text{tail}}$. New items and infrequent queries---which make up a significant fraction of query volume---receive no training signal at all.

\textbf{Manual annotation does not scale.} Human annotation produces clean labels but costs \$0.10--0.50 per pair. At the scale we need (240M+ examples across 4M queries), this would cost \$24M--\$120M per refresh cycle---infeasible for weekly retraining.

\section{Method}
\label{sec:method}

\begin{figure*}[t]
\centering
\begin{tikzpicture}[
    node distance=0.4cm and 0.5cm,
    source/.style={rectangle, draw, rounded corners=2pt, minimum width=1.6cm, minimum height=0.55cm, align=center, font=\scriptsize, fill=blue!8},
    stage/.style={rectangle, draw, rounded corners=3pt, minimum width=2.2cm, minimum height=0.65cm, align=center, font=\small\bfseries, fill=gray!10},
    detail/.style={rectangle, draw, rounded corners=2pt, minimum width=2.2cm, minimum height=0.5cm, align=center, font=\scriptsize, fill=white},
    sample/.style={rectangle, draw, rounded corners=2pt, minimum width=1.5cm, minimum height=0.45cm, align=center, font=\tiny, fill=green!8},
    loss/.style={rectangle, draw, rounded corners=2pt, minimum width=2.0cm, minimum height=0.45cm, align=center, font=\tiny, fill=orange!10},
    arr/.style={-{Stealth[length=2mm]}, thick},
    darr/.style={-{Stealth[length=1.5mm]}, thin, dashed},
    bigarr/.style={-{Stealth[length=3mm]}, very thick, gray!60},
]


\node[stage] (s1) {Stage 1: Collection};
\node[source, above left=0.35cm and -0.2cm of s1] (dict) {Dictionary};
\node[source, above=0.35cm of s1] (bm25) {BM25};
\node[source, above right=0.35cm and -0.2cm of s1] (ann) {ANN (V2)};
\node[detail, below=0.25cm of s1] (s1d) {4M queries $\times$ top-$K$ items + rank positions};

\draw[arr] (dict) -- (s1);
\draw[arr] (bm25) -- (s1);
\draw[arr] (ann) -- (s1);
\draw[darr] (s1) -- (s1d);

\node[stage, right=4.0cm of s1] (s2) {Stage 2: Annotation};
\node[detail, above=0.5cm of s2] (s2u) {Cross-Encoder (184M)\\$\to$ LLM (2B) $\to$ LLM (8B)};
\node[detail, below=0.25cm of s2] (s2d) {89.1\% agreement $\cdot$ 5-class labels $\cdot$ per-class calibration};

\draw[arr] (s1) -- (s2);
\draw[darr] (s2u) -- (s2);
\draw[darr] (s2) -- (s2d);

\coordinate (mid12) at ($(s1)!0.5!(s2) + (0,-2.5)$);


\node[stage, below=3.5cm of s1] (s3) {Stage 3: Extraction};

\node[sample, above left=0.5cm and 0.1cm of s3] (ep) {Easy Pos};
\node[sample, above right=0.5cm and 0.1cm of s3] (hp) {Hard Pos};
\node[sample, below=0.5cm of s3, xshift=-1.7cm] (hn) {Hard Neg};
\node[sample, below=0.5cm of s3, xshift=0cm] (shn) {Token-Sim Neg};
\node[sample, below=0.5cm of s3, xshift=1.7cm] (en) {Random Neg};

\draw[bigarr] (s2d.south) -- ++(0,-0.4) -| (s3.north);
\draw[darr] (s3.north west) ++(0.1,0) -- (ep.south);
\draw[darr] (s3.north east) ++(-0.1,0) -- (hp.south);
\draw[darr] (s3.south) ++(-1.3,0) -- (hn.north);
\draw[darr] (s3.south) -- (shn.north);
\draw[darr] (s3.south) ++(1.3,0) -- (en.north);

\node[stage, right=4.0cm of s3] (s4) {Stage 4: Curriculum};

\node[loss, above=0.5cm of s4] (l1) {S1: BCE --- Easy Pos + Random Neg};
\node[loss, below=0.5cm of s4, xshift=-1.6cm] (l2) {S2: MNR --- Hard Pos/Neg};
\node[loss, below=0.5cm of s4, xshift=1.6cm] (l3) {S3: Triplet --- Token-Sim Neg};

\draw[arr] (s3) -- (s4);
\draw[darr] (s4) -- (l1);
\draw[darr] (s4.south) ++(-0.5,0) -- (l2.north);
\draw[darr] (s4.south) ++(0.5,0) -- (l3.north);
\draw[arr] (l1.south) -- ++(0,-0.15) -| (l2.north west);
\draw[arr] (l2.east) -- (l3.west);

\node[detail, right=1.0cm of s4, fill=yellow!10] (out) {Two-Tower\\BERT $\to$\\ANN Index};
\draw[arr] (s4) -- (out);

\end{tikzpicture}
\caption{Overview of the training-data pipeline. \textbf{Top row:} Stage~1 collects retrieval results with rank positions from three heterogeneous channels; Stage~2 annotates all pairs via a calibrated model cascade (89.1\% agreement with human annotators). \textbf{Bottom row:} Stage~3 classifies pairs into five difficulty levels using channel agreement, rank thresholds, and catalog mining; Stage~4 trains a two-tower BERT model through three curriculum stages of increasing difficulty (BCE $\to$ MNR $\to$ Triplet), with each stage matched to specific sample types. Arrows between training stages indicate checkpoint transfer.}
\label{fig:architecture}
\end{figure*}
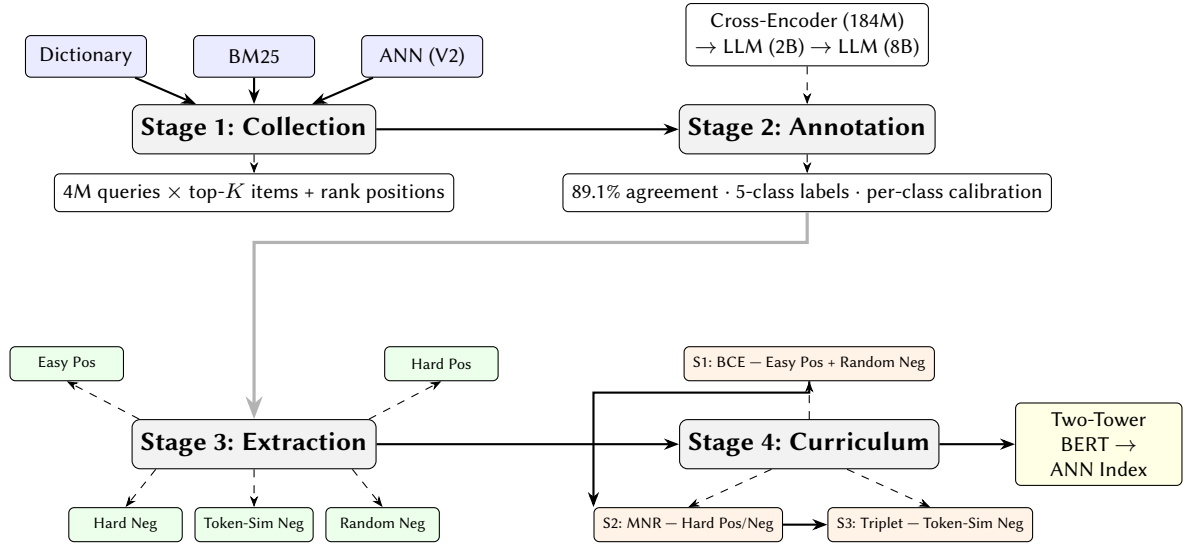

Our pipeline has four stages (Figure~\ref{fig:architecture}): (1) collect retrieval results with rank information from three production channels, (2) annotate all query-item pairs with a calibrated LLM cascade, (3) classify pairs into five difficulty levels using channel agreement and rank thresholds, and (4) train the student model through a three-stage curriculum. We describe each stage below.

\subsection{Retrieval Infrastructure}
\label{sec:infrastructure}

The data collection stage leverages three heterogeneous retrieval channels operating in parallel on Walmart's sponsored search platform, each providing a distinct view of query-item relevance:
\begin{itemize}[leftmargin=*]
    \item \textbf{Dictionary-based text retrieval ($R_D$)}: A BM25-variant retrieval system on the advertising catalog whose item index is augmented with historically relevant queries, enabling exact-match retrieval for known query-item associations and falling back to lexical BM25 scoring otherwise.
    \item \textbf{BM25-based retrieval ($R_B$)}: Standard BM25~\cite{robertson2009probabilistic} retrieval on the advertising catalog using item title, description, brand, and product type as index fields, without query augmentation.
    \item \textbf{ANN retrieval ($R_A$, Embedding Model V2)}: The production neural retrieval model using a two-tower BERT~\cite{devlin2019bert} architecture trained with click-based labels and approximate nearest neighbor search~\cite{wang2024semantic}.
\end{itemize}
For each query $q$, we collect the top-$K{=}500$ retrieved items from each channel, along with each item's rank position $r_s(q, i)$ within channel $s$. These channels exhibit low pairwise overlap---dictionary-ANN overlap is 13.4\% and dictionary-BM25 overlap is 15.05\% at $K{=}500$---meaning the majority of retrieved items exist in disagreement regions. This low overlap is a feature: it provides abundant structured signal for training data extraction (Section~\ref{sec:multichannel}).

\subsection{Calibrated Cascade Annotation}
\label{sec:annotation}

The key idea of our annotation approach is simple: use a cascade of progressively larger fine-tuned classifiers to label query-item pairs with 5-class graded relevance, and calibrate each model so that confident predictions are accepted early while ambiguous pairs are deferred to larger models. The cascade produces hard (argmax) labels, not soft probabilities. This is distant supervision~\cite{mintz2009distant}, not classical knowledge distillation~\cite{hinton2015distilling}: the cascade is used once, offline, and the student is trained with standard supervised objectives. We make this choice deliberately because it decouples the annotator from the student: (1) annotations are computed in batch and cached; (2) the cascade adds no serving latency; and (3) cascade and student can be upgraded independently. The cost is that we discard the cascade's uncertainty on borderline pairs; Section~\ref{sec:discussion} discusses this trade-off.

\textbf{Cascade architecture.} The annotation engine ($f_T$) is a calibrated model cascade that routes query-item pairs through three progressively larger domain-specific fine-tuned classifiers~\cite{chen2023frugalgpt}: (1) a cross-encoder (184M parameters), (2) a LoRA-adapted 2B-parameter LLM, and (3) a LoRA-adapted 8B-parameter LLM. Each model produces a five-class relevance prediction with a calibrated confidence score. If the confidence exceeds a per-model threshold, the prediction is accepted; otherwise, the pair is deferred to the next, larger model. Pairs that remain unresolved after all three models are decided by majority vote with a max-prediction tiebreaker.

\textbf{Per-class isotonic calibration.} A critical design choice is calibrating each predicted class separately via isotonic regression, rather than applying a single global calibration. This captures heterogeneous confidence distributions across classes---models tend to be overconfident on extreme classes (0 and 4) and underconfident on middle classes (1--3). Per-class calibration achieves the best cascade routing accuracy compared to temperature scaling, Platt scaling, and global isotonic methods.

\textbf{Annotation quality.} We audit the cascade against trained human annotators on a held-out set of 100K query-item pairs disjoint from all training data. The cascade reaches 89.1\% agreement on the 5-class task, with the Cross-Encoder handling 74.5\% of pairs at 91.2\% agreement and larger LLMs reserved for ambiguous cases---reducing compute by approximately 50\% relative to running all three models on every pair. For context we also report zero-shot agreement of off-the-shelf GPT-4 (68.2\%) and Claude-3 (70.1\%) on the same audit set under a matched prompt template. We emphasize that this is a \emph{fine-tuned domain cascade vs.\ zero-shot general-purpose LLM} comparison, not a like-for-like evaluation of LLM capability; the relevant takeaway is that off-the-shelf use would inject substantial label noise into the pipeline, motivating our domain-adapted cascade. At the scale of 240M+ examples, the residual $\sim$11\% label noise in cascade labels is tolerable: prior work on noise-robust deep learning~\cite{rolnick2017deep, natarajan2013noisy} shows that deep classifiers converge to near-clean performance when supervised with enough loosely-correlated labels, and our multi-channel agreement filter (Section~\ref{sec:multichannel}) further removes the most likely false negatives.

\subsection{Structured Training Data Extraction}
\label{sec:multichannel}

The key idea of our data mining approach is to use channel agreement and disagreement as a signal for training difficulty. Given the annotated query-item pairs from the cascade, we classify each pair into one of five difficulty levels based on three signals: which channels retrieved the item, where it ranked, and what the cascade labeled it. The pipeline operates over 4 million queries spanning head, torso, and tail segments. Let $\mathcal{R}_D(q)$, $\mathcal{R}_B(q)$, and $\mathcal{R}_A(q)$ denote the retrieval sets from dictionary-based, BM25-based, and ANN retrieval, respectively.

\textbf{Easy Positives ($\mathcal{D}_{\text{EP}}$):} Items retrieved by all three channels and confirmed relevant by the cascade---high-confidence positives where all systems agree:
\begin{equation}
    \mathcal{D}_{\text{EP}} = \{(q, i) : i \in \mathcal{R}_D \cap \mathcal{R}_B \cap \mathcal{R}_A,\; \text{rel}(q, i) \geq 3\}
\end{equation}

\textbf{Hard Positives ($\mathcal{D}_{\text{HP}}$):} Items missed by the ANN retrieval but ranked highly by at least one lexical channel, and confirmed relevant---these are items the current dense model fails to surface. Formally, $(q, i) \in \mathcal{D}_{\text{HP}}$ if $i \in (\mathcal{R}_D(q) \cup \mathcal{R}_B(q)) \setminus \mathcal{R}_A(q)$, there exists $s \in \{D, B\}$ with $r_s(q, i) \leq 50$, and $\text{rel}(q, i) \geq 3$.

\textbf{Hard Negatives ($\mathcal{D}_{\text{HN}}$):} Items ranked within the top 100 by exactly one channel but judged irrelevant. We apply a \emph{multi-channel agreement filter}: negatives retrieved by more than one channel are removed, since cross-channel agreement suggests the item may be borderline relevant despite the cascade label. Formally, $(q, i) \in \mathcal{D}_{\text{HN}}$ if there exists exactly one channel $s \in \{D, B, A\}$ such that $i \in \mathcal{R}_s(q)$ and $r_s(q, i) \leq 100$, and $\text{rel}(q, i) \leq 2$.

\textbf{Semi-Hard (Token-Similar) Negatives ($\mathcal{D}_{\text{SHN}}$):} Items from the catalog with moderate token overlap (TF-IDF similarity $\in [0.2, 0.6]$) but \emph{not retrieved by any channel}, additionally filtered by the cascade to remove false negatives. These lexically plausible but semantically irrelevant examples force the model to learn beyond keyword matching:
\begin{equation}
    \mathcal{D}_{\text{SHN}} = \{(q, i) : \text{TF-IDF}(q, i) \in [0.2, 0.6],\; i \notin \mathcal{R}_D \cup \mathcal{R}_B \cup \mathcal{R}_A,\; \text{rel}(q, i) \leq 2\}
\end{equation}

\textbf{Easy (Random) Negatives ($\mathcal{D}_{\text{EN}}$):} Randomly sampled catalog items with minimal lexical overlap, providing foundational negative signal and broad coverage.

Table~\ref{tab:data_composition} summarizes the composition of the resulting training dataset. After classification, we apply quality controls: (1) removal of queries where no retrieved item receives $\text{rel} \geq 3$ (these queries lack sufficient positive support for contrastive training; approximately 8\% of queries fall into this category, which includes both legitimate no-relevant-ad queries and potential systematic annotation failures), (2) deduplication of near-identical items within the same query-rating group, (3) rank-based filtering (top-50 for positives, top-100 for hard negatives), and (4) per-query sampling limits of 50 positives and 50 negatives. The theoretical maximum of $\sim$400M examples (4M $\times$ 100) is reduced to 240M+ by these filters: tail queries often yield fewer than 50 positives from channel agreement, rank constraints exclude lower-ranked items, deduplication removes near-duplicate product listings, and all-negative queries are discarded entirely. The resulting pipeline produces 240M+ examples (158M+ positives, 80M+ negatives).

\begin{table}[t]
\centering
\caption{Training data composition. Five sample types are extracted across 4M queries, producing 240M+ examples.}
\label{tab:data_composition}
\small
\begin{tabular}{@{}llrr@{}}
\toprule
\textbf{Sample Type} & \textbf{Source} & \textbf{Count} & \textbf{Curriculum} \\
\midrule
Easy Positives & 3-channel agreement & \multirow{2}{*}{158M+} & Stage 1 \\
Hard Positives & Lexical $\setminus$ ANN &  & Stage 2--3 \\
\midrule
Hard Negatives & Single-channel, top-100 & \multirow{3}{*}{80M+} & Stage 2 \\
Token-Sim.\ Negatives & Catalog TF-IDF mining & & Stage 3 \\
Random Negatives & Catalog sampling & & Stage 1 \\
\bottomrule
\end{tabular}
\end{table}

\subsection{Progressive Curriculum Training}
\label{sec:curriculum}

The key idea of our training strategy is to present examples in order of increasing difficulty, so the model learns basic relevance patterns before tackling hard cases. Training on all 240M+ examples at once with a single loss function underperforms a staged approach~\cite{bengio2009curriculum}: hard negatives and token-similar negatives are only useful after the model can distinguish obviously relevant from obviously irrelevant items. We train the two-tower BERT model in three stages, where each stage uses a different loss function matched to the difficulty of its sample types (Table~\ref{tab:data_composition}). Each stage initializes from the best checkpoint of the previous stage.

\textbf{Stage 1: Binary Cross-Entropy (Foundation).} The model learns basic relevance discrimination using only the highest-confidence positives ($\mathcal{D}_{\text{EP}}$, rating = 4, i.e.\ ``Excellent'') and random negatives ($\mathcal{D}_{\text{EN}}$). We deliberately exclude rating-3 (``Good'') items from this stage: including borderline positives alongside trivially-distinguishable random negatives would introduce label ambiguity before the model has learned any relevance signal, harming early convergence. BCE is appropriate here because the positive-negative distinction is unambiguous, requiring only pointwise scoring:
\begin{equation}
    \mathcal{L}_{\text{BCE}} = -\big[y \cdot \log(\sigma(\tau \cdot s(q, i))) + (1-y) \cdot \log(1-\sigma(\tau \cdot s(q, i)))\big]
    \label{eq:bce}
\end{equation}
where $s(q, i) = \cos(E_q(q), E_i(i)) \in [-1, 1]$ is the cosine similarity between the query encoder output $E_q(q)$ and item encoder output $E_i(i)$, and $\tau$ is a learned scalar temperature initialized at 20. The temperature is necessary: without it, $\sigma(s)$ is bounded in $[\sigma(-1), \sigma(1)] \approx [0.27, 0.73]$, preventing the loss from reaching zero for perfect predictions and severely compressing the gradient signal. The temperature-scaled variant is standard in cosine-based contrastive losses~\cite{radford2021clip}.

\textbf{Stage 2: Multiple Negatives Ranking Loss (Discrimination).} The model learns to distinguish relevant items from hard negatives ($\mathcal{D}_{\text{HN}}$) that fooled at least one production system. MNR is chosen over BCE because it explicitly optimizes the ranking objective---the model must score the positive higher than \emph{all} in-batch negatives simultaneously:
\begin{equation}
    \mathcal{L}_{\text{MNR}} = -\log \frac{\exp(s(q, i^+) / \tau)}{\exp(s(q, i^+) / \tau) + \sum_{j=1}^{n} \exp(s(q, i^-_j) / \tau)}
    \label{eq:mnr}
\end{equation}
where $\tau$ is the same learned temperature as in Stage~1 (carried over via checkpoint transfer). This stage combines hard positives ($\mathcal{D}_{\text{HP}}$, rating $\geq 3$) with hard negatives, exposing the model to items the current ANN system retrieves incorrectly. A known limitation of in-batch negative sampling is that another query's positive may be partially relevant to the current query, creating false negatives. We mitigate this through large batch sizes (which reduce per-pair collision probability) and by relying on the subsequent Triplet stage to refine the embedding space; a more principled solution such as debiased contrastive estimation~\cite{robinson2021contrastive} is left for future work.

\textbf{Stage 3: Triplet Loss (Semantic Nuance).} The model learns fine-grained discrimination using token-similar negatives ($\mathcal{D}_{\text{SHN}}$) that share vocabulary with the query but are semantically irrelevant. Triplet loss with margin $\alpha$ directly enforces an embedding-space gap between positive and negative items, which is critical when the negatives are lexically similar and thus close in the initial embedding space:
\begin{equation}
    \mathcal{L}_{\text{triplet}} = \max\big(0,\; d(q, i^+) - d(q, i^-) + \alpha\big)
    \label{eq:triplet}
\end{equation}
where $d(a, b) = 1 - \cos(E(a), E(b))$ is cosine distance (so that lower values indicate higher similarity, consistent with margin semantics) and $\alpha = 0.3$ is the margin hyperparameter selected via grid search on a held-out validation set. This stage forces the model to learn beyond surface-level keyword matching.

\textbf{Model architecture.} The student model is a two-tower architecture based on sentence-transformers~\cite{reimers2019sentencebert}, initialized from a 6-layer, 22.7M-parameter transformer encoder that produces 384-dimensional embeddings. The query and item encoders \emph{share all transformer layers}---a Siamese configuration that reduces model size and regularizes the shared embedding space. The query encoder receives the raw query string; the item encoder receives the product title. No additional product attributes (description, brand, category) are used in the current version; incorporating structured item features is a direction for future work. Training uses AdamW~\cite{loshchilov2019adamw} with learning rate $1 \times 10^{-5}$, 5\% linear warmup, batch size 512 per GPU across 4 GPUs (effective batch size 2048), mixed-precision (float16), and \texttt{torch.compile} optimization. Models are exported to ONNX format and served via a GPU inference server with approximate nearest neighbor search, meeting production latency requirements.

\section{Experimental Setup}
\label{sec:experiments}

\subsection{Evaluation Dataset}

Offline evaluation uses a held-out set of production queries that are \emph{entirely disjoint from the 4M queries used for training data extraction}. No query in the evaluation set appears in any stage of the training pipeline---neither in the multi-channel retrieval collection, the cascade annotation, nor the curriculum training. This separation ensures that offline metrics reflect generalization to unseen queries, not memorization of training examples.

The evaluation queries are stratified by traffic volume. The segment sizes are unequal for practical reasons:

\begin{itemize}[leftmargin=*]
    \item \textbf{Head}: 1,439 queries (highest-frequency searches)---a small set because few queries individually reach head-level traffic, but each is high-impact.
    \item \textbf{Torso}: 21,185 queries (moderate-frequency searches)---the largest segment, reflecting that torso queries are both numerous and individually frequent enough to annotate comprehensively.
    \item \textbf{Tail}: 7,679 queries (low-frequency, long-tail searches)---a sampled subset, because the space of possible tail queries is virtually unbounded and exhaustive coverage is infeasible. We sample tail queries to ensure representation without claiming coverage of the full long-tail distribution.
\end{itemize}

This stratification means the overall metrics in Table~\ref{tab:overall} are weighted toward torso queries (the largest segment), while the per-segment breakdown in Table~\ref{tab:segments} isolates performance on each frequency tier---particularly tail queries, where click-based models are most deficient.

The evaluation set comprises 30,303 unique queries (1,439 + 21,185 + 7,679). For each query, we evaluate the top-500 retrieved items from each system, producing approximately 15.2 million query-item pairs. To ensure a fair comparison on identical query distributions, we restrict evaluation to queries for which both systems return results; in practice, fewer than 2\% of queries are unique to one system, so this restriction has negligible effect on the comparison.

\textbf{Evaluation labels.} Relevance labels for evaluation are provided by trained \emph{third-party human annotators}---distinct from and independent of the LLM cascade used to produce training labels. Annotators follow the same 5-class graded-relevance guidelines used in Walmart's production quality measurement, and each query-item pair is labeled by two annotators with a third used for adjudication on disagreements; inter-annotator agreement (Cohen's $\kappa$) on a 10K-pair audit is 0.79 (substantial agreement on the 5-class scale), with per-class agreement highest on the extreme ratings (Embarrassing: $\kappa{=}0.88$, Excellent: $\kappa{=}0.95$) and lowest on adjacent middle classes (Okay: $\kappa{=}0.62$, Bad: $\kappa{=}0.68$).
Using human labels for the evaluation set means that the offline metrics in Section~\ref{sec:results} are not subject to the circularity concern of scoring a model against the same supervision signal that trained it: the training set uses cascade labels, but the evaluation set is fully independent human supervision. The 89.1\% cascade-human agreement reported in Section~\ref{sec:annotation} is computed on this same human-labeled pool, providing a consistent calibration point across training and evaluation. The online A/B test (Section~\ref{sec:abtest}) provides a second, annotation-free source of validation through business metrics.

\subsection{Baselines}

We compare Embedding Model V3 (our system) against two baselines on the same 30K human-labeled query set:
\begin{itemize}[leftmargin=*]
    \item \textbf{Embedding Model V2 (primary baseline)}: the previous-generation two-tower BERT retrieval model deployed on Walmart ANN search~\cite{wang2024semantic}, trained using click-based labels with a single-stage training procedure. This is the direct head-to-head comparison: same architecture, same serving stack, different training supervision.
    \item \textbf{Pre-trained base model (reference)}: the same transformer encoder~\cite{reimers2019sentencebert} that both V2 and V3 are initialized from, evaluated without any fine-tuning. This baseline isolates the effect of fine-tuning: any gap between the base model and V2 reflects what click-based training adds, and any gap between V2 and V3 reflects what our LLM-annotated pipeline adds on top.
\end{itemize}
The three retrieval channels (Section~\ref{sec:infrastructure}) are data sources for training extraction, not evaluation baselines.

\subsection{Significance Testing}

For each offline metric we report paired bootstrap 95\% confidence intervals over the 30K evaluation queries (10,000 resamples), and we test V2-vs-V3 differences with a two-sided paired bootstrap test at $\alpha = 0.05$. For the online A/B test, all reported lifts are verified as statistically significant via the platform's internal experimentation framework. All offline V3-vs-V2 differences reported in Section~\ref{sec:results} are significant at $p < 10^{-3}$; 95\% CIs are reported in Table~\ref{tab:overall}.

\subsection{Metrics}

Our primary offline metric is NDCG@10~\cite{jarvelin2002cumulated}, computed with raw 5-class relevance gains (0--4) as the gain function. We also report Precision (fraction of top-10 items with $\text{rel} \geq 3$), MAP, MRR, and Average Relevance (mean score on the 0--4 scale). Recall requires special care in our setting: since the full set of relevant items in the catalog is unknown, we compute pool-based recall over the judged pool (union of top-500 retrieved items from both systems for each query). This may underestimate absolute recall for both systems equally, but relative comparisons remain valid.

\subsection{Online A/B Test Setup}
\label{sec:abtest}

The online evaluation was conducted on Walmart's sponsored search platform over a two-week period with randomized user-bucket assignment:
\begin{itemize}[leftmargin=*]
    \item \textbf{Control}: Embedding Model V2 (production baseline).
    \item \textbf{Treatment}: Embedding Model V3 (our system).
    \item \textbf{Metrics}: Ad spend, CTR, eCPM, ROAS, click conversion rate, revenue per click.
\end{itemize}
Each arm received tens of millions of ad requests over the two-week test period, providing high statistical power for detecting the observed effect sizes.

\section{Results}
\label{sec:results}

We present our results in five parts. First, we report overall offline metrics. Second, we analyze the relevance distribution shift. Third, we break down results by query segment and product category. Fourth, we present online A/B test results. Fifth, we report ablation studies.

\subsection{Offline Evaluation}

Table~\ref{tab:overall} presents the overall offline evaluation results. We include the pre-trained base model (the same encoder that V2 and V3 are initialized from, without any fine-tuning) to isolate the contribution of training. The base model achieves NDCG@10 of 0.771, showing that V2's click-based training already provides a substantial +13.9\% gain (0.771 $\to$ 0.878). Our LLM-annotated pipeline pushes this further to 0.923 (+5.1\% over V2), indicating that V3's improvements are relative to an already well-trained production system.

\begin{table}[t]
\centering
\caption{Overall offline evaluation results (30,303 human-labeled evaluation queries). The pre-trained base model is the same encoder that V2 and V3 are initialized from, evaluated without fine-tuning. Values in brackets are paired bootstrap 95\% CIs on the absolute $\Delta$ (V3 vs.\ V2). All V3-vs-V2 differences are significant at $p < 10^{-4}$ (paired bootstrap, 10K resamples).}
\label{tab:overall}
\small
\begin{tabular}{@{}lcccc@{}}
\toprule
\textbf{Metric} & \textbf{Pre-trained base} & \textbf{Model V2} & \textbf{Model V3} & \textbf{V3 vs.\ V2 (95\% CI)} \\
\midrule
Avg Relevance & 2.040 & 2.413 & 2.687 & +0.274 [+0.264, +0.285] \\
NDCG@10       & 0.771 & 0.878 & 0.923 & +0.045 [+0.042, +0.051] \\
Precision     & 0.413 & 0.526 & 0.624 & +0.098 [+0.093, +0.113] \\
Recall        & 0.509 & 0.676 & 0.801 & +0.125 [+0.121, +0.135] \\
MAP           & 0.588 & 0.722 & 0.820 & +0.098 [+0.095, +0.102] \\
MRR           & 0.811 & 0.901 & 0.945 & +0.044 [+0.038, +0.048] \\
\bottomrule
\end{tabular}
\end{table}

\subsection{Relevance Distribution Shift}

Beyond aggregate metrics, Table~\ref{tab:distribution} reveals the qualitative shift in retrieval quality. Embedding Model V3 reduces the proportion of Embarrassing results (rating 0) from 8.7\% to 3.5\%---a 5.2 percentage point reduction---while increasing Good and Excellent results from a combined 52.7\% to 62.4\%. This shift suggests that the LLM-annotated supervision improves average relevance in part by reducing the most harmful retrieval errors.

\begin{table}[t]
\centering
\caption{Relevance distribution shift. Values are percentages of all retrieved items at each rating level. The largest absolute change is the 5.2pp reduction in embarrassing results.}
\label{tab:distribution}
\small
\begin{tabular}{@{}lrrr@{}}
\toprule
\textbf{Rating} & \textbf{Model V2} & \textbf{Model V3} & \textbf{Change} \\
\midrule
0 -- Embarrassing & 8.7\% & 3.5\% & $-5.2$pp \\
1 -- Bad & 26.6\% & 22.6\% & $-4.0$pp \\
2 -- Okay & 12.0\% & 11.5\% & $-0.5$pp \\
3 -- Good & 20.0\% & 26.6\% & $+6.6$pp \\
4 -- Excellent & 32.7\% & 35.8\% & $+3.1$pp \\
\bottomrule
\end{tabular}
\end{table}

\begin{figure}[t]
\centering
\begin{tikzpicture}
\begin{axis}[
    ybar,
    bar width=7pt,
    width=\columnwidth,
    height=5.5cm,
    ylabel={NDCG@10},
    symbolic x coords={Head,Torso,Tail,Overall},
    xtick=data,
    ymin=0.78, ymax=1.0,
    legend style={at={(0.5,1.02)}, anchor=south, legend columns=2, font=\small},
    every axis plot/.append style={fill opacity=0.85},
    enlarge x limits=0.2,
    grid=major,
    grid style={dashed, gray!30},
    nodes near coords,
    nodes near coords style={font=\tiny, rotate=90, anchor=west},
    every node near coord/.append style={/pgf/number format/.cd, fixed, precision=3},
]
\addplot coordinates {(Head,0.913) (Torso,0.893) (Tail,0.830) (Overall,0.878)};
\addplot coordinates {(Head,0.959) (Torso,0.934) (Tail,0.886) (Overall,0.923)};
\legend{Model V2, Model V3}
\end{axis}
\end{tikzpicture}
\caption{NDCG@10 by query segment. Tail queries show the largest relative improvement (+6.8\%), consistent with the hypothesis that LLM-annotated supervision addresses the data-sparsity challenge inherent to click-based training.}
\label{fig:segment_comparison}
\end{figure}
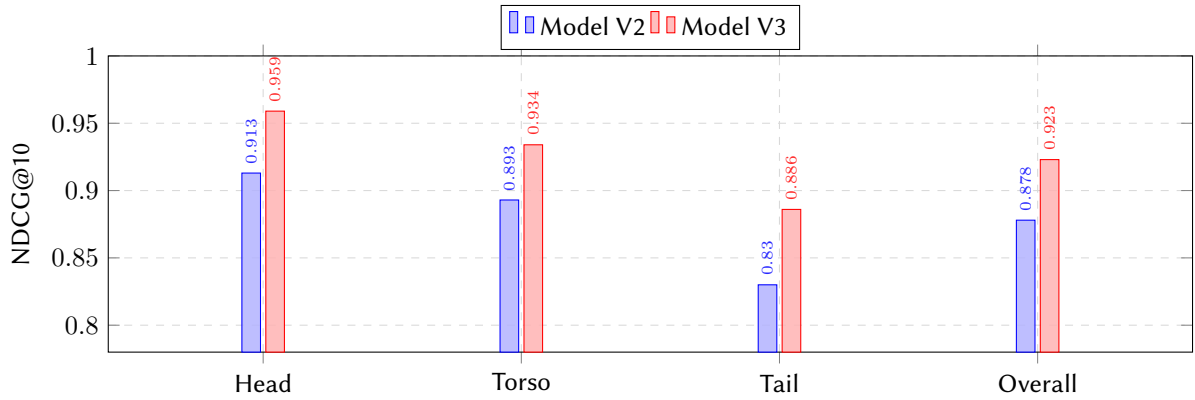

Figure~\ref{fig:segment_comparison} compares NDCG@10 and precision across query segments, highlighting that improvements are universal but largest for tail queries where click-based training is most deficient.

\subsection{Performance by Query Segment}

Table~\ref{tab:segments} shows that improvements are consistent across all query segments, with particularly notable gains for tail queries where the production baseline performs poorest. Tail queries improve from 0.830 to 0.886 NDCG@10 (+6.8\%), the largest relative gain among all segments. This is consistent with our hypothesis: cascade annotations do not depend on historical click volume, so tail queries receive the same annotation quality as head queries, closing the data-sparsity gap that degrades click-trained models.

\begin{table}[t]
\centering
\caption{Performance by query segment (evaluation queries). NDCG@10 and Avg Relevance reported.}
\label{tab:segments}
\small
\begin{tabular}{@{}lrcccc@{}}
\toprule
\textbf{Segment} & \textbf{Queries} & \multicolumn{2}{c}{\textbf{NDCG@10}} & \multicolumn{2}{c}{\textbf{Avg Relevance}} \\
\cmidrule(lr){3-4} \cmidrule(lr){5-6}
 & & V2 & V3 ($\Delta$) & V2 & V3 ($\Delta$) \\
\midrule
Head & 1,439 & 0.913 & 0.959 (+5.1\%) & 2.527 & 2.812 (+11.3\%) \\
Torso & 21,185 & 0.893 & 0.934 (+4.6\%) & 2.469 & 2.733 (+10.7\%) \\
Tail & 7,679 & 0.830 & 0.886 (+6.8\%) & 2.239 & 2.536 (+13.3\%) \\
\bottomrule
\end{tabular}
\end{table}

\subsection{Performance by Product Category}

Table~\ref{tab:categories} shows consistent improvements across all six major product categories, from high-performing categories (Apparel: +3.1\%) to challenging ones (Food: +7.1\%, from 0.837 to 0.897 NDCG@10). The consistency of improvements across diverse product categories suggests that the multi-channel training data and LLM annotation generalize well across product domains.

\begin{table}[t]
\centering
\caption{Performance by product category (evaluation queries, NDCG@10).}
\label{tab:categories}
\small
\begin{tabular}{@{}lccc@{}}
\toprule
\textbf{Category} & \textbf{Model V2} & \textbf{Model V3} & \textbf{Improvement} \\
\midrule
Apparel & 0.933 & 0.962 & +0.029 (+3.1\%) \\
Consumables & 0.894 & 0.935 & +0.041 (+4.5\%) \\
Entertainment & 0.883 & 0.923 & +0.040 (+4.5\%) \\
Food & 0.837 & 0.897 & +0.060 (+7.1\%) \\
Hardlines & 0.905 & 0.946 & +0.041 (+4.5\%) \\
Home & 0.925 & 0.957 & +0.032 (+3.4\%) \\
\bottomrule
\end{tabular}
\end{table}

\subsection{Online A/B Test Results}

Table~\ref{tab:abtest} presents the results of the two-week online A/B test conducted on Walmart's sponsored search platform. With tens of millions of ad requests per arm, the test has high statistical power for the observed effect sizes. Embedding Model V3 achieves consistent improvements across all business metrics, providing annotation-independent evidence that the offline relevance gains translate to real economic value.

\begin{table}[t]
\centering
\caption{Online A/B test results (two weeks, tens of millions of ad requests per arm, randomized user-bucket assignment). Relative lift of Embedding Model V3 (treatment) over Embedding Model V2 (control). All lifts are statistically significant as verified by the platform's experimentation framework.}
\label{tab:abtest}
\small
\begin{tabular}{@{}lc@{}}
\toprule
\textbf{Metric} & \textbf{Relative Lift} \\
\midrule
Ad Spend & +2.80\%$^{*}$ \\
Ad CTR & +1.4\%$^{*}$ \\
eCPM & +2.8\%$^{*}$ \\
ROAS Direct & +2.4\%$^{*}$ \\
Clicks Conv.\ Direct & +2.9\%$^{*}$ \\
Rev/Click Direct & +3.9\%$^{*}$ \\
\bottomrule
\multicolumn{2}{@{}l@{}}{\footnotesize $^{*}$ Statistically significant ($p < 0.05$).}
\end{tabular}
\end{table}

The simultaneous improvement in both efficiency metrics (eCPM, CTR) and downstream business metrics (ROAS, conversion, revenue per click) is consistent with a positive feedback loop: more relevant ads may attract more clicks, which improves advertiser ROI and can incentivize higher bids, though we cannot isolate these causal mechanisms from a single A/B test.

\subsection{Component Analysis and Ablations}
\label{sec:ablations}

We isolate the contribution of each major component through a set of controlled ablations on the full human-labeled evaluation set. Table~\ref{tab:ablations} reports NDCG@10 for each ablation, with V3 (full system) at the top as the reference and individual component removals grouped into three categories: training objective, training data mining, and annotation cascade. We discuss each group in turn.

\begin{table}[t]
\centering
\caption{Ablation study (30,303 human-labeled evaluation queries). Each row removes or modifies one component from the full V3 system (NDCG@10 = 0.923).}
\label{tab:ablations}
\small
\begin{tabular}{@{}lcc@{}}
\toprule
\textbf{Configuration} & \textbf{NDCG@10} & \textbf{$\Delta$ vs.\ V3} \\
\midrule
V3 (full: BCE $\to$ MNR $\to$ Triplet) & 0.923 & --- \\
\midrule
\multicolumn{3}{@{}l@{}}{\emph{Training objective (curriculum ordering)}} \\
\quad -- single-stage (all data mixed) & 0.843 & $-0.080$ ($-8.7\%$) \\
\quad -- MNR $\to$ Triplet (drop BCE) & 0.903 & $-0.020$ ($-2.2\%$) \\
\quad -- Triplet $\to$ MNR (drop BCE) & 0.897 & $-0.026$ ($-2.8\%$) \\
\quad -- BCE $\to$ MNR (drop Triplet) & 0.883 & $-0.040$ ($-4.3\%$) \\
\quad -- BCE $\to$ Triplet (drop MNR) & 0.857 & $-0.066$ ($-7.2\%$) \\
\quad -- MNR $\to$ BCE (drop Triplet) & 0.843 & $-0.080$ ($-8.7\%$) \\
\quad -- Triplet $\to$ BCE (drop MNR) & 0.844 & $-0.079$ ($-8.6\%$) \\
\midrule
\multicolumn{3}{@{}l@{}}{\emph{Training data mining}} \\
\quad -- single-channel extraction (ANN only) & 0.875 & $-0.048$ ($-5.2\%$) \\
\quad -- without token-similar (semi-hard) negatives & 0.886 & $-0.037$ ($-4.0\%$) \\
\quad -- without multi-channel agreement filter & 0.916 & $-0.007$ ($-0.8\%$) \\
\quad -- without hard positives (lexical $\setminus$ ANN) & 0.915 & $-0.008$ ($-0.9\%$) \\
\midrule
\multicolumn{3}{@{}l@{}}{\emph{Annotation cascade}} \\
\quad -- Platt scaling instead of isotonic & 0.904 & $-0.019$ ($-2.1\%$) \\
\quad -- cross-encoder only (no LLM fallback) & 0.914 & $-0.009$ ($-1.0\%$) \\
\quad -- global isotonic calibration (no per-class) & 0.917 & $-0.006$ ($-0.7\%$) \\
\quad -- 8B LLM only (no routing, $\sim$3$\times$ cost) & 0.921 & $-0.002$ ($-0.2\%$) \\
\bottomrule
\end{tabular}
\end{table}

\textbf{Curriculum training.} The three-stage curriculum is the single largest contributor: single-stage training on the same 240M examples scores 0.843, while the full BCE $\to$ MNR $\to$ Triplet curriculum reaches 0.923 (+9.5\%). The six two-stage variants reveal two clear patterns. First, \emph{stage ordering matters enormously}: any configuration ending with BCE collapses to near-single-stage performance (MNR $\to$ BCE = 0.843, Triplet $\to$ BCE = 0.844). BCE is a foundation loss that teaches coarse relevance discrimination; using it as a refinement stage after a ranking loss destroys the fine-grained structure the ranking loss learned. Second, \emph{MNR before Triplet is better than Triplet before MNR}: MNR $\to$ Triplet = 0.903 vs.\ Triplet $\to$ MNR = 0.897. The listwise ranking objective (MNR) provides a stronger intermediate representation for the pairwise margin objective (Triplet) to refine. The BCE foundation stage adds +2.2\% on top of the best two-stage variant (0.903 $\to$ 0.923), confirming that the coarse-to-fine progression is genuinely additive rather than redundant.

\textbf{Training data mining.} Multi-channel extraction is the largest data-side contributor: restricting to a single channel (ANN only) costs 5.2\% NDCG@10 (0.875 vs.\ 0.923). Token-similar negatives are the second largest at 4.0\% (0.886), confirming that Stage 3's lexically-plausible distractors are essential for teaching the model to look beyond surface keyword matching. The agreement filter and hard positives contribute smaller but consistent gains (0.8\% and 0.9\% respectively). The agreement filter's modest impact suggests that cross-channel agreement is a useful but not critical signal for negative quality; the bulk of the data mining value comes from channel diversity and the token-similarity mining.

\textbf{Annotation cascade.} The cascade ablations show that calibration method matters more than model selection. Platt scaling is the worst performer (0.904, $-2.1\%$), substantially worse than per-class isotonic (0.923) and global isotonic (0.917, $-0.7\%$). This confirms the value of per-class calibration for heterogeneous confidence distributions. On the model side, using only the cross-encoder costs 1.0\% but handles 74.5\% of pairs, while using the 8B LLM alone scores 0.921---nearly matching the full cascade at roughly 3$\times$ the compute cost. The cascade's value is primarily economic: it achieves equivalent accuracy at a fraction of the annotation budget.

\textbf{Summary.} Ordered by impact: (1) three-stage curriculum with correct ordering (+9.5\%), (2) multi-channel extraction (+5.2\%), (3) token-similar negatives (+4.0\%), (4) per-class isotonic calibration (+2.1\% vs.\ Platt), (5) cross-encoder-to-LLM cascade routing (+1.0\%), (6) hard positives (+0.9\%), (7) agreement filter (+0.8\%), and (8) per-class vs.\ global isotonic (+0.7\%). The first three components account for the majority of the improvement, which aligns with the paper's framing: the structured data mining and curriculum design are the core contributions.

\section{Discussion}
\label{sec:discussion}

We discuss why our approach works, what trade-offs it involves, and where it still falls short.

\textbf{Why does LLM-annotated supervision work?} We believe three factors contribute. First, the cascade provides relevance signals that do not depend on click position or query frequency. Unlike click-trained models, which receive no signal for tail queries, the cascade judges every query-item pair on text semantics alone. The cascade has its own biases---inherited from its fine-tuning data---but these biases are systematic and predictable, unlike the heterogeneous noise in click signals. Second, the multi-channel mining pipeline creates training examples at five difficulty levels, so the model sees a diverse range of failure modes: missed relevant items, retrieved irrelevant items, and lexically similar distractors. Third, the curriculum prevents the model from being overwhelmed by hard examples before it learns basic relevance patterns, which the +9.5\% NDCG@10 gain over single-stage training confirms empirically. The ablation results further reveal that stage ordering is critical: ending with the coarse BCE loss destroys the fine-grained structure learned by ranking losses, collapsing performance to near-single-stage levels.

\textbf{Why multi-channel disagreement matters.} The low overlap between retrieval channels (13.4\% dictionary-ANN, 15.05\% dictionary-BM25 at $K{=}500$) is a feature, not a limitation. It means most items fall into regions where systems disagree, and these are exactly the cases that matter most for training. Agreement regions provide high-confidence positives. Disagreement regions provide hard positives (items lexical systems find but the ANN misses) and hard negatives (items one system retrieves incorrectly). These are qualitatively different from random negatives: they are the items that confuse real production systems, making them maximally informative for contrastive training.

\textbf{Why tail queries improve the most.} The largest relative gain is on tail queries (+6.8\% NDCG@10), where click-trained models perform worst due to sparse engagement data. This is expected: the cascade assesses relevance based on text, so rare queries and new items receive the same annotation quality as popular ones. Combined with the multi-channel mining that surfaces hard positives for tail queries, our approach closes the quality gap between query frequency segments.

\textbf{Hard labels vs.\ soft labels: a trade-off.} We use hard (argmax) cascade labels rather than soft probability distributions. This decouples the cascade from the student entirely: annotations are computed once, cached, and reused across training runs and model architectures. However, it discards the cascade's uncertainty on borderline pairs (ratings 1--3), where collapsing the distribution to a single class may propagate systematic biases. Quantifying this gap through a controlled comparison of hard-label training against soft-label distillation from the same cascade is the most informative follow-up to this work.

\textbf{Where does the model still fail?} The Food category shows the lowest absolute NDCG@10 (0.897) even after improvement. Food queries often involve ambiguous specifications (e.g., ``organic milk'' vs.\ specific brands) and high synonym variation. This suggests that the current title-only input could benefit from structured product attributes such as brand, category, and description. The token-similar negative mining may also be less effective for categories where genuinely different products share most of their vocabulary.

\subsection{Limitations}

We note the following limitations of our work. (a) \emph{Proprietary data}: our training and evaluation data are proprietary to Walmart and cannot be publicly released, though the methodology is described in sufficient detail to be reproduced on other corpora. (b) \emph{Training-label noise}: the 89.1\% cascade-human agreement means roughly 11\% of training labels diverge from human judgment, with the largest errors on borderline classes (1--3). (c) \emph{Comparison breadth}: our primary comparison is against the production V2 baseline; a broader comparison against fine-tuned general-purpose retrievers would further characterize where gains come from. (d) \emph{Multi-channel dependency}: the pipeline requires multiple heterogeneous retrieval systems; platforms with a single channel would need to create synthetic disagreement. (e) \emph{Single deployment}: the online A/B results come from one two-week test; replication on additional markets and longer time windows would strengthen the evidence.

\subsection{Production Considerations}

The system runs in production with a weekly retraining cadence. The fully automated pipeline---from multi-channel data extraction through cascade annotation, curriculum training, and ONNX export---eliminates manual annotation and enables continuous model improvement. Serving uses GPU-accelerated inference with approximate nearest neighbor search, meeting production latency requirements. The total retraining cycle completes within 24--48 hours on GPU infrastructure.

A subtlety of weekly retraining is the feedback loop: the ANN channel ($R_A$) is the current production model, so the hard positives shift each week as the ANN improves. In practice, we observe that this loop is stable---each iteration's hard-positive set shrinks as the ANN closes gaps with the lexical systems. However, we have not formally characterized the convergence properties of this loop, and pathological drift remains a theoretical risk.

\section{Conclusion}
\label{sec:conclusion}

How can we train dense retrieval models at production scale without click signals or manual annotation? In this paper, we answered this question with an end-to-end pipeline that combines multi-channel retrieval mining, calibrated LLM cascade annotation (89.1\% agreement with human annotators), structured sample extraction across five difficulty levels, and three-stage progressive curriculum training. The pipeline generates 240M+ training examples across 4M queries and enables fully automated weekly model retraining.

We deployed the resulting model on Walmart's sponsored search and evaluated it against 30K queries with independent third-party human labels. The system improves NDCG@10 by +5.1\% over the click-trained production baseline, with gains across all query segments and product categories. Notably, the largest gain appears on tail queries (+6.8\%), where click-based models have the least training signal. Embarrassing retrievals drop by more than half (8.7\% $\to$ 3.5\%). A two-week online A/B test with tens of millions of ad requests per arm provides independent validation: +2.80\% ad spend, +2.8\% eCPM, and +2.9\% click conversion rate.

Overall, our work provides a practical blueprint for replacing click-based training with structured LLM-annotated supervision in production retrieval systems. The key insight---that multi-channel retrieval disagreement is a rich source of structured training signal---is applicable to any platform with multiple retrieval systems. Future work will pursue three directions: (1) enriching the input representation with structured product attributes and multimodal signals, as the current title-only approach underperforms on visually-driven and attribute-heavy categories; (2) comparing hard-label distant supervision against soft-label distillation from the same cascade to quantify the information loss from label discretization; and (3) extending the framework to cross-market and multilingual settings.

\section*{Declaration on Generative AI}
During the preparation of this work, the author(s) used Generative AI only for formatting assistance in plotting code used to produce figures, and not for drafting, analysis, interpretation, or any other part of the manuscript. The author(s) reviewed, corrected, and validated the generated code and figures and take(s) full responsibility for the publication's content.

\bibliography{references}

\end{document}